\documentclass[preprint]{aastex}

\begin{document}

\title{The sdB+M Eclipsing System HW Virginis and its Circumbinary Planets}
\author{Jae Woo Lee$^{1}$, Seung-Lee Kim$^1$, Chun-Hwey Kim$^2$, Robert H. Koch$^3$, Chung-Uk Lee$^1$, Ho-Il Kim$^1$, and Jang-Ho Park$^{1,2}$}
\affil{$^1$Korea Astronomy and Space Science Institute, Daejeon 305-348, Korea}
\email{jwlee@kasi.re.kr, slkim@kasi.re.kr, leecu@kasi.re.kr, hikim@kasi.re.kr, pooh107162@kasi.re.kr}
\affil{$^2$Department of Astronomy and Space Science, College of Natural Science and Institute
        for Basic Science Research, Chungbuk National University, Cheongju 361-763, Korea}
\email{kimch$@$chungbuk.ac.kr}
\affil{$^3$Department of Physics and Astronomy, University of Pennsylvania, Philadelphia, USA}
\email{rhkoch@earthlink.net}

\begin{abstract}
For the very short-period sdB eclipsing binary HW Vir, we present new CCD photometry made from 2000 through 2008.
In order to obtain consistency of the binary parameters, our new light curves, showing sharp eclipses and
a striking reflection effect, were analyzed simultaneously with previously published radial-velocity data.
The secondary star parameters of $M_2$=0.14 M$_\odot$, $R_2$=0.18 R$_\odot$, and $T_2$=3,084 K are consistent
with those of an M6-7 main sequence star. A credibility issue regarding bolometric corrections is emphasized.
More than 250 times of minimum light, including our 41 timings and spanning more than 24 yrs, were used for
a period study. From a detailed analysis of the $O$--$C$ diagram, it emerged that the orbital period of HW Vir
has varied as a combination of a downward-opening parabola and two sinusoidal variations, with cycle lengths
of $P_3$=15.8 yr and $P_4$=9.1 yr and semi-amplitudes of $K_3$=77 s and $K_4$=23 s, respectively.
The continuous period decrease with a rate of $-$8.28$\times$10$^{-9}$ d yr$^{-1}$ may be produced by
angular momentum loss due to magnetic stellar wind braking but not by gravitational radiation. Of the possible causes
of the cyclical components of the period change, apsidal motion and magnetic period modulation can be ruled out.
The most reasonable explanation of both cyclical variations is a pair of light-travel-time effects driven
by the presence of two substellar companions with projected masses of $M_3 \sin i_3$=19.2 M$\rm_{Jup}$ and
$M_4 \sin i_4$=8.5 M$\rm_{Jup}$. The two objects are the first circumbinary planets known to have been formed
in a protoplanetary disk as well the first ones discovered by using the eclipse-timing method. The detection
implies that planets could be common around binary stars just as are planets around single stars and
demonstrates that planetary systems formed in a circumbinary disk can survive over long time scales. Depending
on the thermal inertia of their massive atmospheres, the hemispheres of the planets turned toward the stars
can experience substantial reciprocating temperature changes during the minutes-long primary eclipse intervals.
\end{abstract}

{}

\section{INTRODUCTION}

With the continuing discovery of extrasolar planets\footnote {http://exoplanet.eu/} and an expectation that the majority
of solar-type stars reside in binary or multiple systems (Duquennoy \& Mayor 1991), planetary formation in binary systems
has become an important matter. Recent theoretical studies (Moriwaki \& Nakagawa 2004; Quintana \& Lissauer 2006;
Pierens \& Nelson 2008b) have predicted that circumbinary planets (i.e., those orbiting two stars which are tightly
bound gravitationally) can form and survive over long time scales. Characterization of such planets is potentially of
great interest because they should provide constraints and information on planet formation and evolution which are
possibly different from those for planets orbiting single stars. There is currently known only one planet orbiting
a binary system, the millisecond pulsar PSR B1620-26 with a white dwarf companion in the globular cluster M4
(Backer et al. 1993). However, Sigurdsson et al. (2003) showed that this planet may not have formed in
a circumbinary disk but have been captured from a passing star in the dense environment of the cluster core. Therefore,
no planet formed in a circumbinary disk around a gravitationally-bound binary has yet been detected unambiguously.

Since the first discovery of planets around the pulsar PSR1257+12 (Wolszczan \& Frail 1992), almost 300 extrasolar planets
have been detected by using radial velocity, photometric transit, microlensing, and direct imaging techniques.
Among all of them, the recently reported multiple-planet system OGLE-2006-BLG-109L is noteworthy in that the mass ratio,
separation ratio, and equilibrium temperatures of the planets are remarkably similar to those of Jupiter and Saturn
(Gaudi et al. 2008). This agrees with theoretical predictions that other systems can resemble the Solar System and
suggests that Solar System analogues may be prevalent. In addition to the methods mentioned above, 5 planets have
been discovered by light-travel-time (hereafter LTT) effects: 4 planets around two pulsars (Backer et al. 1993;
Wolszczan 1994) and one around a pulsating star (Silvotti et al. 2007).

The LTT effect in eclipsing binaries occurs because the distance from a binary to the observer varies due to scaled reflex
motions from substellar companions moving around the barycenter of a multiple system. The mechanism produces variations
of eclipse timings of a binary system, which timings then act as an accurate clock for detecting the other objects.
For their own interest, timings have been exploited since the late 18th century, and now they can be seen to be tools that
potentially offer discovery of companions to many close binary stars. Because of lengthy histories of period monitoring,
the eclipse timing method has the additional advantage of detecting substellar companions to close double stars with
much longer periods than is permitted by the other techniques already mentioned. With timing accuracies of about $\pm$10 s
for selected eclipsing binaries showing sharp eclipses, it should be possible to detect circumbinary planets of
$\sim$10 M$\rm_{Jup}$ in long-period orbits of 10-20 yr around eclipsing systems (Ribas 2006). One such candidate was
recently reported for the eclipsing pair CM Dra (Deeg et al. 2008) but, due to the limited data set, it is not clear
at present whether the third object is a planet or not (Ofir 2008).

\section{HISTORY OF HW VIRGINIS}

HW Vir (BD-07$^{\rm o}$3477, HIP 62157, GSC 5528-0629) was first classified as a subdwarf B (sdB) star by
Berger \& Frigant (1980) and was discovered to be a very short-period eclipsing variable with a period of about 2.8 hours,
$T_1$=26,000 K and $T_2$=4,500 K, and relative polar radii (in units of the separation) of $r_1$=0.203 and $r_2$=0.207
by Menzies \& Marang (1986). Since then, the binary system has been the topic of several investigations and
its observational history was reviewed in a recent paper by \. Ibanoglu et al. (2004). From optical spectroscopy
covering $\Delta$$\lambda$=3704$-$8667 \AA, Wood \& Saffer (1999) determined the effective temperature, gravity,
and helium abundance of the sdB star to be $T_1$=28,488 K, $\log g$=5.63, and $N$(He)/$N$(H)=0.0066, all of which
are typical for a normal sdB star. They also reported weak H$\alpha$ absorption features which they interpreted
as arising from a reflection effect, the result of irradiation of the face of the secondary star closest to the primary.
Their results indicate a secondary star orbital velocity of about 275 km s$^{-1}$. Consequently, they also derived
the stars' physical parameters to be $M_1$=0.48 M$_\odot$, $M_2$=0.14 M$_\odot$, $R_1$=0.176 R$_\odot$, and
$R_2$=0.180 R$_\odot$, with the primary's radius formally a bit smaller the secondary's. Within their errors,
the values of the secondary's mass, radius, and temperature are consistent with a normal main-sequence M star.

After the orbital period changes of HW Vir were examined for the first time by Kilkenny et al. (1994), the period
was studied by \c Cakirl \& Devlen (1999), Wood \& Saffer, Kiss et al. (2000), Kilkenny et al. (2000, 2003), and
\. Ibanoglu et al. Kilkenny et al. (1994) found that the period of the system was decreasing over a 9-yr baseline
and discussed possible causes of the effect: gravitational radiation, rotation of a line of apsides, mass transfer,
a third body in the system, and angular momentum loss (hereafter AML) via magnetic braking in a modest stellar wind.
They suggested the last two mechanisms to be the most plausible causes of the period diminution. The LTT effect was
analyzed in detail by \c Cakirl \& Devlen, who presented third-body parameters with a period of about 19 yr and a mass
of $M_3 \sin i_3$=0.022 M$_\odot$, indicating a substellar companion for any inclination greater than 17$^\circ$.
On the other hand, Wood \& Saffer and Kiss et al. represented the $O$--$C$ residuals by two intersecting linear fits,
corresponding to a sudden period jump between two constant periods around 1991. More recently and from the analysis
of eclipsing timings spanning 17 and 19 yrs, respectively, Kilkenny et al. (2003) and \. Ibanoglu et al. concluded
that the cause of the period change is an LTT effect due to a third body with periods of 20.7 and 18.8 yr and
minimum masses of 0.028 and 0.022 M$_\odot$ in the same order as the two sets of authors.

In 2000 two authors (JWL and C-HK) of this paper reviewed the history of HW Vir and believed that it would be possible to resolve
the confusion associated with the period variability. Consequently, they initiated a long-term photometric monitoring of
the system. In this paper, we first present the results of this newer observing program and light and velocity syntheses
and a period analysis of the system. We then describe the discovery of two circumbinary planets revolving around
the close binary star by using the LTT technique.

\section{NEW LONG-TERM CCD PHOTOMETRY}

New CCD photometric observations of HW Vir were obtained between 2000 and 2008, except
for 2001 and 2006, using the 61-cm reflector at Sobaeksan Optical Astronomy Observatory (SOAO) in Korea. The observations
of the 2000 season were carried out with a $VR$ filter set and a PM512 CCD camera, which has 512$\times$512 pixels and
a field of view (FOV) of about 4$\arcmin$.3$\times$4\arcmin.3. GSC 5528-0761 (C$_1$=BD-08$^{\rm o}$3411; Sp. type=A3)
was chosen as a comparison star during this season since it had been used in the previous observations
by G\" urol \& Selan (1994), \c Cakirl \& Devlen and \. Ibanoglu et al. No check star was observed. Since the FOV was
not large enough to observe the two stars simultaneously on the same CCD frame, we had to alternate between the variable
and comparison stars and the observing duty cycle was not longer than 3 min, corresponding to about 0.018$P$.
The reduction method for the observed frames has been described by Park (1993). The 2000 light curves, together with
those of the other seasons, are shown in Figure 1.

The observations of the other seasons were made with a $V$ filter and a SITe 2K CCD camera, which has
2048$\times$2048 pixels and a FOV of about 20$\arcmin$.5$\times$20$\arcmin.5$. The instrument and reduction method
have been described by Lee et al. (2007). A nearby star, GSC 5528-0655 (C$_2$=BD-07$^{\rm o}$3470), imaged
on the chip at the same time as the program target, was selected as a comparison star, and no peculiar light variations
were detected against measurements of a nearby check star, GSC 5528-0479 (C$_3$). The 1$\sigma$-value of
an individual (C$_3$-C$_2$) magnitude difference is $\pm$0.012 mag. Over all observing seasons, 1897 individual observations
were obtained in the two bandpasses (1687 in $V$ and 210 in $R$) and these are listed in Table 1. The newer light curves
of HW Vir are plotted in the upper panel of Figure 2 as magnitude differences between the variable and the C$_2$ star
{\it versus} orbital phase. The (C$_1$-C$_2$) and (C$_3$-C$_2$) differences are shown in the middle and lower panels,
respectively, wherein we can see that the C$_1$ star is the main source of the scatter seen in Figure 1. In order to
examine the possible intrinsic variability of this star, we applied a multiple-period analysis (Kim \& Lee 1996)
to these magnitude differences. This scrutiny indicated that the C$_1$ star is a multiperiodic variable with
a dominant period of 0.2333 d. Because $\delta$ Scuti variables typically have spectral types A$-$F,
short pulsation periods less than 0.3 d and low amplitudes less than 0.1 mag (Breger 1979), the C$_1$ star
GSC 5528-0761 may be a $\delta$ Scuti-type pulsating star.

In addition to the SOAO observations, eclipse timings were made in both 2006 and 2008 without any filter
at the campus station of the Chungbuk National University Observatory (CbNUO). The CbNUO station functions with
a 35-cm reflector equipped with a SBIG ST-8 CCD camera electronically cooled. The details of the CbNUO observations
have been well described elsewhere (Kim et al. 2006).

\section{LIGHT AND VELOCITY SOLUTIONS}

As can be seen in Figures 1 and 2, the light curve of HW Vir displays sharp eclipses and two light maxima
around phases 0.44 and 0.56.  This strongly indicates a very prominent reflection effect and
a secondary star that likely contributes very little light to the system. There are no significant differences
among the last six seasonal data sets. In order to obtain a consistent light and velocity solution and thereafter
the absolute dimensions of HW Vir, we simultaneously solved our photometric data and
the available radial-velocity curves using modes 0 and 2 of the Wilson-Devinney synthesis code
(Wilson \& Devinney 1971, hereafter WD). In the case of mode 0, both the luminosity ($L_2$) and temperature ($T_2$)
of the secondary star can be adjusted as free parameters, while $L_2$ in mode 2 is coupled to and computed
from its temperature. First of all, we analyzed the HW Vir curves by applying mode 2 but failed to derive
a satisfactory solution, especially around secondary eclipse. In this mode the temperature of the cool star
came out significantly too hot (more than 20\%) for its mass and radius derived by previous workers and
subsequently by us too. This effect apparently was caused because the cool star contributes such a feeble fraction
of systemic light and because of the large flux seated in the reflection effect. These modelling defects indicated
that mode 2 is inadequate for modelling HW Vir and so we chose mode 0 instead.

In this analysis, the effective temperature ($T_1$) of the sdB star was assumed to be 28,488$\pm$208 K given
by Wood \& Saffer. The linear bolometric ($X_{1,2}$) and monochromatic ($x_{1,2}$) limb-darkening coefficients
were interpolated from the values of van Hamme (1993) and used in concert with the model atmosphere option.
The gravity-darkening exponents were assumed to be $g_1$=1.0 and $g_2$=0.32, while the bolometric albedos were
initialized at $A_1$=$A_2$=1.0 because of the strong reflection effect seated in the the secondary star. Also,
synchronous rotation for both components was assumed, so that $F_{1}$=$F_{2}$=1. Thus, adjustable parameters are
the system velocity ($\gamma$), the semi-major axis ($a$), the mass ratio ($q$), the orbital inclination ($i$),
the temperature ($T_2$) of the secondary star, the dimensionless surface potentials ($\Omega_{1,2}$) and
the monochromatic luminosities ($L_{1,2}$) of both components. Initial values for all these parameters were
taken from previous studies. Here, the subscripts 1 and 2 refer to the sdB star and its companion, respectively.
Throughout this analysis, the method of multiple subsets (Wilson \& Biermann 1976) was used.

To weight reasonably the observations in a simultaneous velocity and light solution, we used a weighting scheme similar
to that for the eclipsing binary RU UMi (Lee et al. 2008; cf. Wilson 1979). Table 2 lists the velocity and light curves
of HW Vir analyzed in this paper and their standard deviations ($\sigma$) used to assign data-dependent weights.
The radial-velocity curve of Menzies \& Marang was not studied because their measurements have not yet been released.
Weights inversely proportional to the square root of the light level were applied
to the light measures and individual points were employed during the simultaneous fittings with the WD code.
Because the RV1 of Hilditch et al. (1996) has a scatter smaller than the RV1 of Wood \& Saffer, we adopted
a $\sigma$ value of 3.39 km s$^{-1}$ for the former in order to generate the curve-dependent weight factors
for the combined RV1 data.

Our analysis of the HW Vir curves was carried out in two stages. In the first stage, the parameters mentioned above
were the only ones adjusted. In the second stage, after we had tested the albedo of the secondary star from 1.0
to 0.4, $A_2$, $x_1$, and $x_2$ were included as additional free variables. The resulting binary parameters appear
as Model 1 in Table 3. The synthetic radial velocity curves are shown in Figure 3, while the residuals of
the (post-2000) observations from the model light curves are plotted in the upper panel of Figure 4. These plots show that
our binary model fits both kinds of observations quite well.  In our representation, the primary eclipse is
a partial transit and our results resemble closely those of previous investigators in a non-dimensioned sense.
There is, however, a difference from former interpretations which called for a partial occultation at primary eclipse.
This discrepancy illustrates the well-known difficulty of distinguishing a transit from an occultation if the radii
of the two stars are very similar and eclipses are partial. Because the observational weight of our light curves
is much higher than that of any previous one, we prefer our interpretation. With these results in hand, we turned
to the 2000-season light curves treating them in the same way. The results, shown in Table 4
(where the footnotes used in Table 3 apply as well) and the light residuals appearing in the lower panels of Figure 4,
are completely consistent with those for the later seasons, implying stability of the light curve.  Naturally,
the parameter errors for the 2000 season are larger because the light curve noise is greater.

Even though the weight of our light curves is much higher than those of previous ones, several second-order photometric
parameters did not emerge with very precise evaluations. The light contribution from the secondary star being so small,
there was never any hope of discovering its limb or gravity darkening coefficients with high precision. This is borne
out by the values and errors in Tables 3 and 4. The albedo of the primary star would be poorly determined as well
and so was left at the assumed value.

There is a major concern about the solution results because of the independent sets of radial velocities. Although taken
in the same observing season and each obtained over only a short interval, the two data sets show a large discrepancy
of about 11 km s$^{-1}$ in systemic velocity. The measurement (+1.82 km s$^{-1}$) of Wood \& Saffer is much more
positive than that (-9.17 km s$^{-1}$) of Hilditch et al. In neither paper is there a hint of a problem with
wavelength assignments but Wood \& Saffer assign a remarkably large error of 21.9 km s$^{-1}$ to the mean velocity of
their rest frame.  If the given error is not a typo, the systemic velocity from the other study might be preferred.
In order to see how the discrepancy propagates into the other system parameters, we changed
the Wood \& Saffer velocity zero point to agree with that of Hilditch et al. and ran the WD code again. These results
appear as Model 2 in Table 3 and it can be seen that the differences between the two models and the changes in
formal errors are very small.  The seeming reason for the small changes is that the {\it ad hoc} change in
the systemic velocity is only a 2\% fraction of the peak-to-peak velocity amplitude. The Wood \& Saffer velocities
also bear on another analytical detail. Tables 3 and 4 show that the stars are very nearly spheres so there is
little observational and analytical leverage in determining $q$ from the light curve and RV1 measures alone.
The values of $q$ given in the tables are, however, reasonably consistent with the value that can be generated from the
weak absorption features that Wood \& Saffer found after considerable manipulation of their spectra.

Because our orbital period study discussed in the following section indicates a continuous period decrease, we asked
the WD code to examine the time derivative of the period and obtained a fractional rate of $-$(1.8$\pm$0.2)$\times$10$^{-11}$
and, hence, $-$(6.6$\pm$0.7)$\times$10$^{-9}$ d yr$^{-1}$. Within about 2$\sigma$, this will be seen to agree with the value
$-$(8.28$\pm$0.06)$\times$10$^{-9}$ d yr$^{-1}$ derived from the eclipse timings themselves.

\section{THE EVOLUTION OF HW VIR IN THE MILKY WAY GALAXY}

The difficulty associated with the radial velocity curves has almost no impact on calculated absolute stellar parameters
and these are given in Table 5, where the radii ($R$) are the mean-volume radii evaluated using the tabulations of
Mochnacki (1984). Our gravity value is identical, within errors, to that which Wood \& Saffer obtained spectroscopically
and the mean densities are acceptable for the two stars. The luminosity ($L$) and the bolometric magnitudes ($M_{\rm bol}$)
were obtained by adopting $T_{\rm eff}$$_\odot$=5,780 K and $M_{\rm bol}$$_\odot$=+4.69 (Popper 1980).
The bolometric corrections (BCs) were obtained from the relation between $\log T$ and BC recalculated from the table
of Flower (1996) by Kang et al. (2007). Using an apparent visual magnitude of $V$=+10.5 at maximum light and ignoring
interstellar reddening at the galactic coordinates of $\ell$=299$^\circ.9$ and $b$=54$^\circ.2$, we calculate a distance
to the system of 181$\pm$20 pc. This is small compared to 546 pc from the useless trigonometric parallax (1.83$\pm$1.94 mas)
from the {\it Hipparcos} and {\it Tycho} Catalogues (ESA 1997), while it is consistent with 171$\pm$19 pc computed
by Wood \& Saffer. The radius of the secondary star is in excellent accord with that (0.17 R$_\odot$) obtained by using
an empirical mass-radius relation for low-mass main-sequence stars (Bayless \& Orosz 2006) and its parameters would
correspond to a spectral type of approximately M6-7. HW Vir is therefore a detached eclipsing binary consisting of
a sdB primary and a main-sequence M companion.

There is a major discrepancy between Tables 3 and 5. The light ratio of Table 3 translates into an $M_{\rm V}$ magnitude difference
of +8.81 whereas that from Table 5 is +11.37.  This difference of 2.56 is not caused by a typo in the light ratio from
the light curves nor is it due to an incorrect temperature ratio which enters the two separate calculations
in the same way. Our opinion is that the discrepancy is caused by the calibration of the bolometric corrections as
a function of temperature. Had, for instance, we used Popper's (1980) or Cox's (2000) calibrations, the changed entries
of Table 5 would diminish the difference to +1.02 and +0.82, respectively. We prefer the light ratio generated by the WD code
and the magnitude difference calculated from it.

We construe the evolution of this binary to have proceeded within the familiar theoretical frameworks for single stars and
close binaries. In general, sdB stars, also known as extreme horizontal branch stars and found in both the old disk
and halo populations, are considered to be core-helium-burning objects with core masses of about 0.5 M$_\odot$ and
hydrogen envelopes that are too thin ($\le$ 0.02 M$_\odot$) to sustain nuclear burning (Heber 1986; Saffer et al. 1994).
When HW Vir was newly on the main sequence, the present sdB object may be supposed to have had a mass comparable to
Sun's.  Nuclear evolution drove it to the red-giant state, which state had a significant effect on the binary configuration,
for the cool star had to have been enveloped in the expanding envelope of the red giant. During this common envelope stage,
systemic mass and angular momentum losses led to the diminution of the system's orbital size and period bringing it to
its present detached condition. This description is not significantly different from that of Wood \& Saffer.

Hot subdwarfs should evolve directly to the white dwarf cooling sequence after core-helium exhaustion, bypassing
the asymptotic giant branch and planetary nebula intervals. If the progenitor of the sdB component is assumed to have
had a ZAMS mass of 1 M$_\odot$, the present age of HW Vir is at most 12 Gyr (Han et al. 2002; Hu et al. 2007).
The current secondary component may be considered unevolved. According to canonical stellar evolution theory for stars
with masses below 0.30 M$_\odot$, this star should be fully convective. The object will attain a contact state via
Roche-lobe contraction due to loss of orbital angular momentum rather than endure a radius increase due to
nuclear evolution (Ritter 1986). The orbital period at which the current secondary will come into contact with
its Roche lobe is about 0.073 d at which point the binary starts its existence as a semi-detached system. This evolution
of HW Vir beyond its present state will be driven by AML through magnetic braking caused by stellar winds and eventually
by gravitational radiation.  The time to achieve Roche lobe contact and hence to initiate
mass transfer is about 1.4 Gyr, which is longer than the time-scale ($\sim 10^{8}$ yr) that the sdB primary star
needs to evolve into a white dwarf (Wood et al. 1993; Hu et al.). The sdB star will become a white dwarf before mass transfer
begins and thus before HW Vir ultimately becomes a cataclysmic variable (CV). At that time, the white dwarf will accrete
material from the secondary star so that the binary system will endure violent and powerful mass outflows.

One implication of the sdB star having begun as only a solar-mass object and evolved to its present state is that HW Vir
is an old system. With an accurate photometric parallax and good {\it Hipparcos} proper motions, it is possible to
calculate the components of its space velocity: the object is descending from the north toward the galactic plane with
a galactocentric velocity of about 8.5 km s$^{-1}$ and must be near apogalacticum. It is appropriate to consider it
a member of the thick-disk population and to have passed through a relatively low-stellar-density volume of the Milky Way.
There is, therefore, a reasonable expectation that any companions to the binary are equally old and have not been captured
from some other former parent.

\section{THE $O$--$C$ VARIATION AND ITS IMPLICATIONS}

From all our CCD observations, times of minimum light have been determined using the method of Kwee \& van Woerden (1956).
There are 41 new timings given in Table 6, of which the first eight are weighted means from the observations in the $V$ and
$R$ bandpasses. In this table, the means of the errors for the SOAO primary minima were calculated to be 0.00006 d in 2000
and 0.00003 d in the other seasons. This difference arose because of two reasons:
(1) the comparison star used in 2000 was variable and (2) the $V$- and $R$-filter observations are significantly
less frequently sampled than are the single $V$-filter data. For a system with the very short period of 2.8 h,
this sampling difference contributes to a worse precision of the minimum timing from weighted two-filter measures than
from single-filter observations.

For the period study of HW Vir, we have collected 217 timings (164 photoelectric and 53 CCD) from the literature to add
to our measurements. Most of the eclipse timings were extracted from the database published by Kreiner et al. (2001).
To construct the $O$--$C$ diagram of HW Vir, we first used the linear terms of the \. Ibanoglu et al. ephemeris:
\begin{equation}
  C = \mbox{HJD}~2,445,730.55743 + 0.1167195E.
\end{equation}
The resulting $O$--$C$ residuals with regard to this ephemeris are plotted in the upper panel of Figure 5, where CC
and PE code for CCD and photoelectric minima, respectively, and I and II indicate the primary and secondary eclipses,
respectively.

First of all, we examined whether the period variations of HW Vir could be represented by a single-cyclical ephemeris
as several previous researchers (\c Cakirl \& Devlen; Kilkenny et al.; Ibanoglu et al.) have suggested. Fitting
all timing residuals to that ephemeris form failed to give a satisfactory result. After testing several other forms,
such as a quadratic {\it plus} a single-cyclical ephemeris and a two-cyclical ephemeris, we found that
the $O$--$C$ variation is best fitted by the combination of a downward-opening parabola and two cyclical variations.
The sinusoidal terms were provisionally identified as LTT effects. Thus, the eclipse timing residuals were finally
fitted to a quadratic {\it plus} two-LTT ephemeris (Lee et al. 2007):
\begin{eqnarray}
  C = T_0 + PE + AE^2 + \tau_{3} + \tau_{4},
\end{eqnarray}
where $\tau_{3}$ and $\tau_{4}$ are the LTTs due to two additional bodies (Irwin 1952) and each includes
five parameters ($a_{12}\sin i$, $e$, $\omega$, $n$, $T$). Here, $a_{12}\sin i$, $e$ and $\omega$ are
the orbital parameters of the eclipsing pair around the mass center of the quadruple system. The parameters $n$ and
$T$ denote Keplerian mean motion of the mass center of the eclipsing pair and the epoch of its periastron passage,
respectively. In this fitting process, because many timings were published without timing errors, we calculated
the standard deviations of all residuals in order to assign mean errors for each observational method and type of
eclipse as follows: $\pm$0.00010 for CCD I, $\pm$0.00015 for CCD II, $\pm$0.00008 for PE I, and $\pm$0.00017 for PE II.
Weights for the timings were then scaled as the inverse squares of these values.

The Levenberg-Marquardt algorithm (Press et al. 1992) was applied to solve for the thirteen coefficients of
the quadratic {\it plus} two-LTT ephemeris (Irwin 1959). The result is plotted in Figure 5 and summarized in Table 7,
together with related quantities. The absolute parameters of HW Vir, listed in Table 5, have been used for this and
subsequent calculations. From the detailed analysis of the $O$--$C$ diagram, we have therefore found that
the orbital period of HW Vir has varied with two sinusoidal components superimposed on a downward-opening parabola.

The quadratic term ($A$) of equation (2) represents a continuous period decrease with a rate of
d$P$/d$t$ = $-$8.3$\times$10$^{-9}$ d yr$^{-1}$, which may be due to AML from the binary system.
The period derivative corresponds to angular momentum ($J$) loss at a rate of d$J$/d$t$ = $-$3.7$\times$10$^{35}$
in cgs units. AML in a pre-CV such as HW Vir could be explained by two possible mechanisms:
gravitational radiation or magnetic braking. The AML rate due to gravitational radiation is given by
\begin{equation}
\biggl({{\rm d}J \over {\rm d}t} \biggr)_{\rm gr} = -3.47 \times 10^{-67} {(M_{1}M_{2})^2 \over (M_{1}+M_{2})^{2/3}} P^{-7/3},
\end{equation}
where $M_1$ and $M_2$ represent the masses of the eclipsing primary and secondary, respectively, and all quantities
are expressed in cgs units (Paczynski 1967). With HW Vir parameters listed in Table 5, we get a theoretical rate of
$-$1.0$\times$10$^{33}$ (again in cgs units) for the supposed gravitational radiation, which is two orders of magnitude
too small to be the single cause of the secular period change.

The other mechanism is AML caused by magnetic stellar wind braking in the secondary star, which must have
a deep convective atmosphere. The standard model for magnetic braking in CVs was developed by extrapolation
from studies of braking rates for solar-type stars in open clusters. Rappaport et al. (1983) established
the following empirical relationship that is commonly used in CV studies:
\begin{equation}
  \biggl({{\rm d}J \over {\rm d}t} \biggr)_{\rm mb} = -3.8 \times 10^{-30} M_{2} {R_{\odot}}^4 \biggl({R_{2} \over R_{\odot}}\biggr)^{\gamma} \omega_{2}^3,
\end{equation}
where $\gamma$ is a dimensionless parameter in the range from 0 to 4 and $\omega_{2}$ is the angular rotation velocity.
For 0 $\le \gamma \le$ 4, an AML rate between $-6.1\times$10$^{36}$ and $-5.7\times$10$^{33}$ is obtained, which limits
are consistent with the observed AML rate. This means that AML due to magnetic braking as used in standard models of
close binary evolution can explain the secular period decrease of HW Vir satisfactorily and a value of the order of 2
might be appropriate for $\gamma$.

We now examine whether the cyclical components of the period variability can be explained uniquely or not.
Apsidal motion is not a viable explanation because the orbital eccentricity of the eclipsing system is negligible
and the $O$--$C$ residuals for the primary and secondary eclipses change in phase and amplitude with each other.
There remain only the possibilities of a magnetic activity cycle in the cool secondary star, or LTT effects due to
other bodies physically linked to the eclipsing pair, or some combination of these two causes. Firstly,
the period variations could be produced by magnetic period modulations seated in the M-type secondary star,
as was initially proposed by Applegate (1992) and later modified by Lanza et al. (1998). With the periods ($P_{3,4}$)
and amplitudes ($K_{3,4}$ ) listed in Table 7 and with the absolute dimensions of HW Vir, the model parameters
for each cyclical component were calculated from the Applegate formulae and are listed in Table 8. According to
the Applegate parameters, the secondary component with 0.003 L$_\odot$ should exhibit rms luminosity variations
of 0.035 L$_\odot$ and 0.017 L$_\odot$ for the long- and short-term period variations, respectively. Quite apart
from the numerical impossibility of modulations of such amplitudes, changes of this magnitude were never observed
in our light curves between 2000 and 2008. Therefore, an Applegate mechanism cannot even contribute to
the sinusoidal variations. Secondly, as can be seen in Figure 5, all observed timings agree quite well with
light-time effects imposed by third and fourth bodies. We calculated periods of $P_3$=15.8 yr and $P_4$=9.1 yr, and
semi-amplitudes of $K_3$=77 s and $K_4$=23 s for the third and fourth components, respectively. It is worth noting
that the LTT value for the orbital dimension of the eclipsing orbit itself is somewhat smaller than 3 s.
According to Kepler's third law in a multiple system, the amplitudes and time scales of eclipse timing variations
depend on both the masses and periods of the additional objects. Thus, the mass functions of the 3rd and 4th objects
become $f_3 (M)$=1.5$\times10^{-5}$ M$_\odot$ and $f_4 (M)$=1.2$\times10^{-6}$ M$_\odot$, which imply minimum masses
of 0.0184 M$_\odot$ and 0.0081 M$_\odot$ in the same order. The two LTT orbits have relatively high determinacy
because the observations already cover about 1.5 and 3 cycles for the long and short periods, respectively.

\section{SUMMARY AND DISCUSSION}

We have argued that the most obvious explanation of the sinusoidal variations of the eclipse timing residuals are LTT effects
by two substellar companions with masses of $M_3 \sin i_3$=19.2$\pm$0.2 M$\rm_{Jup}$ and $M_4 \sin i_4$=8.5$\pm$0.4 M$\rm_{Jup}$,
where M$\rm_{Jup}$=M$_\odot$/1047. Simulations (e.g., Bonnell \& Bate 1994; Holman \& Wiegert 1999) show that circumbinary planets
(a) can be formed by the binary interaction with its circumbinary disk, (b) have stable orbits in all binary configurations,
and (c) are most likely coplanar with their host binaries. So, it is not unreasonable that the two substellar companions
have orbits nearly coplanar with HW Vir itself in its orbital inclination of about 81$^\circ$ as seen from Earth.

When the timing residuals of the eclipsing pair reach an extremum for a given sinusoid of the ephemeris, that
companion must be at either inferior or superior conjunction. A meager light change, due to either a transit across
the eclipsing stars or an occultation of the companion by those stars, is possible for appropriate inclinations.
We looked for these effects among the residuals from our light curve model but found no convincing light variations.
At the separations of the companions from the stars,
orbits co-planar with that of the eclipsing binary would forbid transits and occultations at the conjunctions.

From a theoretical argument based upon mass, brown dwarfs (BDs) should be born with masses between the least massive stars
of $\sim$75 M$\rm_{Jup}$ and the most massive planets of $\sim$13 M$\rm_{Jup}$ (Burrows et al. 2001). However,
observational results (Grether \& Lineweaver 2006; Udry \& Santos 2007) indicate that the BD and planet populations overlap
in the 10-20 M$\rm_{Jup}$ interval so it is not possible to differentiate between low-mass BDs and high-mass giant planets from
measured masses alone. Additional information on the formation and evolution of such substellar objects is needed. It is
likely that BDs form by fragmentation of a protostellar cloud while planets are built up from protoplanetary disk material
(Udry \& Santos). Below the BD desert around 31 M$\rm_{Jup}$ (Grether \& Lineweaver), the companion with $a$=5.3 AU,
$e$=0.46 and mass of 19.2 M$\rm_{Jup}$ is best regarded as a planet formed simultaneously with the other companion with
$a$=3.6 AU, $e$=0.31 and mass of 8.5 M$\rm_{Jup}$ in a circumbinary disk. Distances of the two planets from
the parent binary and their orbital periods are similar to those of the asteroid belt-Jupiter system in the Solar System.
The periods suggest nearly 5:3 or 2:1 resonant captures. This fact and the relatively high eccentricities are in line
with theoretical results on planet-planet interaction and stability (Kley et al. 2004; Ford \& Rasio 2008;
Pierens \& Nelson 2008a) and support a two-planet explanation. Finally, this detection of two giant planets demonstrates
that planetary systems formed in a circumbinary disk around two stars can survive over long time scales.

From physical parameters of HW Vir, the binary-planet separations, and assumed Bond albedoes of 0.343 (just as for Jupiter),
we estimate the effective temperatures of the more and less massive planets to be ~230 K and ~270 K, respectively -
thus about two times that of Jupiter. These values are dominated by the flux from the primary star. When that object
is nearly totally eclipsed every 2.8 hours, the equilibrium temperatures at the tops of the star-directed planetary hemispheres
should drop by a factor of about 9 over 10.4 minutes and then recover to the quoted values over a similar time interval.
Any resulting changing weather patterns depend on the thermal inertia and structures of these massive atmospheres.

The {\it Spitzer Space Telescope} recently observed debris disks orbiting both members, rather than just one star, of
various main-sequence binary systems (Trilling et al. 2007)). This means that circumbinary planets around binary stars
could be just as common as planets around single stars. The two giant planets around HW Vir may be expected to
have formed in a similar circumbinary disk. According to the theory of migration and evolution of planets embedded
in such a disk, these two planets grew as they sustained stable orbits. The stellar pair HW Vir has evolved into
a close system from an initial wide binary due to AML. After the massive component evolved through
the red-giant phase, the binary star passed into its current configuration while losing very large amounts of
mass (almost half of the original amount) during its common-envelope ejection episode (Han et al. 2002, 2003).
It is certain that the masses of both planets increased by accreting material during that red-giant expansion
of the presently hot star and the common envelope ejection. It is even possible that the current secondary star
was able to accrete enough mass to become a core-hydrogen-burning star after a previous existence as a BD.
As in the case of the sdB star V391 Peg (Silvotti et al 2007), the two planets orbiting HW Vir may have migrated
outward due to the reduced mass of the binary system and still have survived around the close pair. We know
nothing about the fates of the two planets when HW Vir becomes a CV, but the existence of the PSR B1620-26-system
suggests that they are likely to survive bound to the binary.

The discovery and interpretation of planetary systems orbiting binaries such as HW Vir will help in understanding
the diverse architectures of extrasolar planetary systems and especially of planets around evolved stars.
With dedicated observation, theoretical modeling and numerical simulation, the evolution of a close binary and
the formation and evolution of its planetary system can potentially be amalgamated within a unified understanding.

\acknowledgments{ }
We would like to thank the staff of the Sobaeksan Optical Astronomy Observatory for assistance with our observations.
We appreciate the careful reading and valuable comments of the anonymous referee.
This research has made use of the Simbad database maintained at CDS, Strasbourg, France. C-HK acknowledges support
by a Korea Research Foundation Grant funded by the Korean Government (MOEHRD, Basic Research Promotion Fund, KRF-2005-015-C00188).

\clearpage
\begin{figure}
 \includegraphics[]{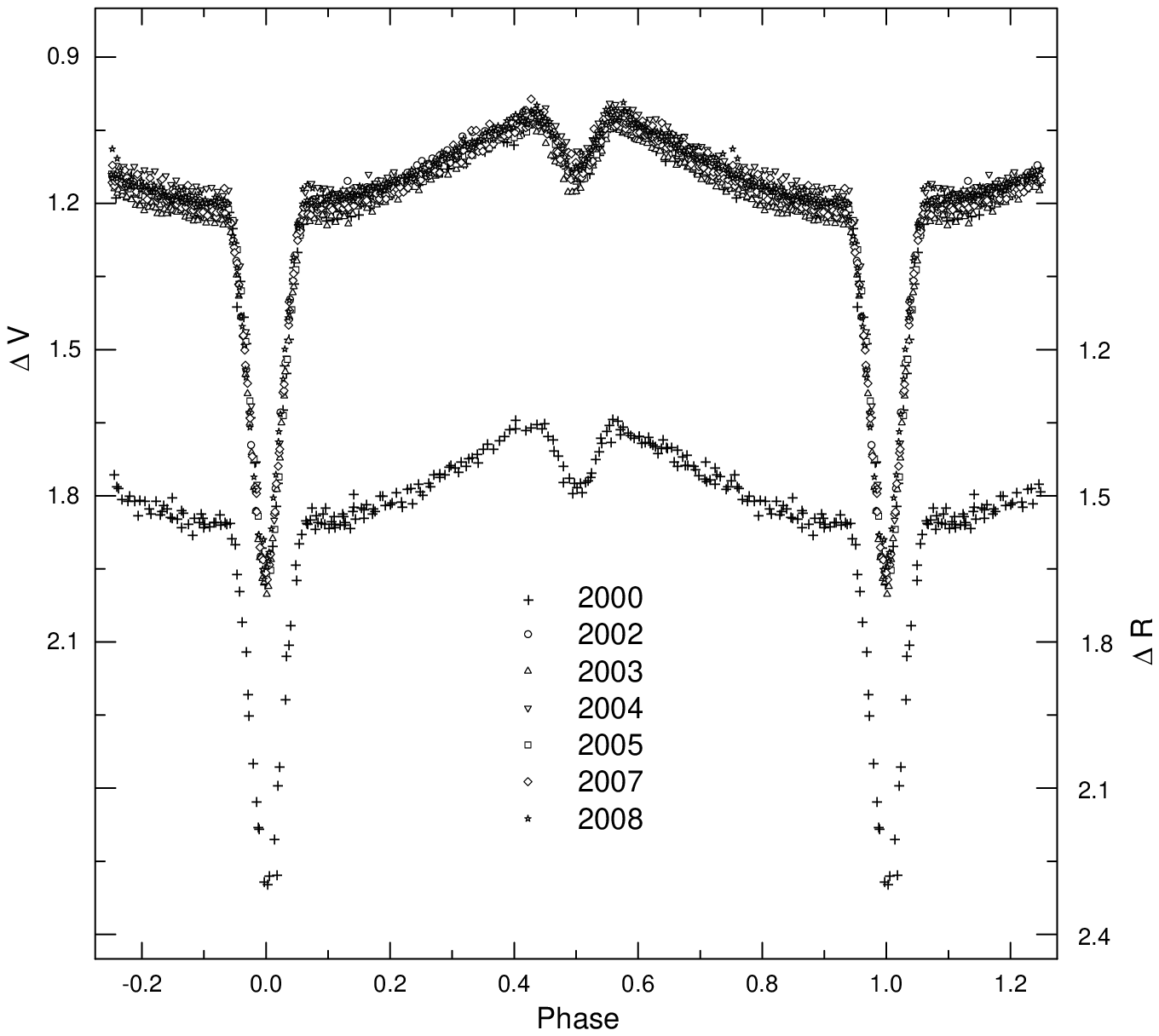}
 \caption{Light curves of HW Vir in the $V$ and $R$ bandpasses. All measures are referred to the C$_1$ star
  used in many previous observations. The observed peak-to-peak noise is about four times larger than
  the uncertainties ($\pm$0.01 mag) of the observations.}
 \label{Fig1}
\end{figure}

\begin{figure}
 \includegraphics[]{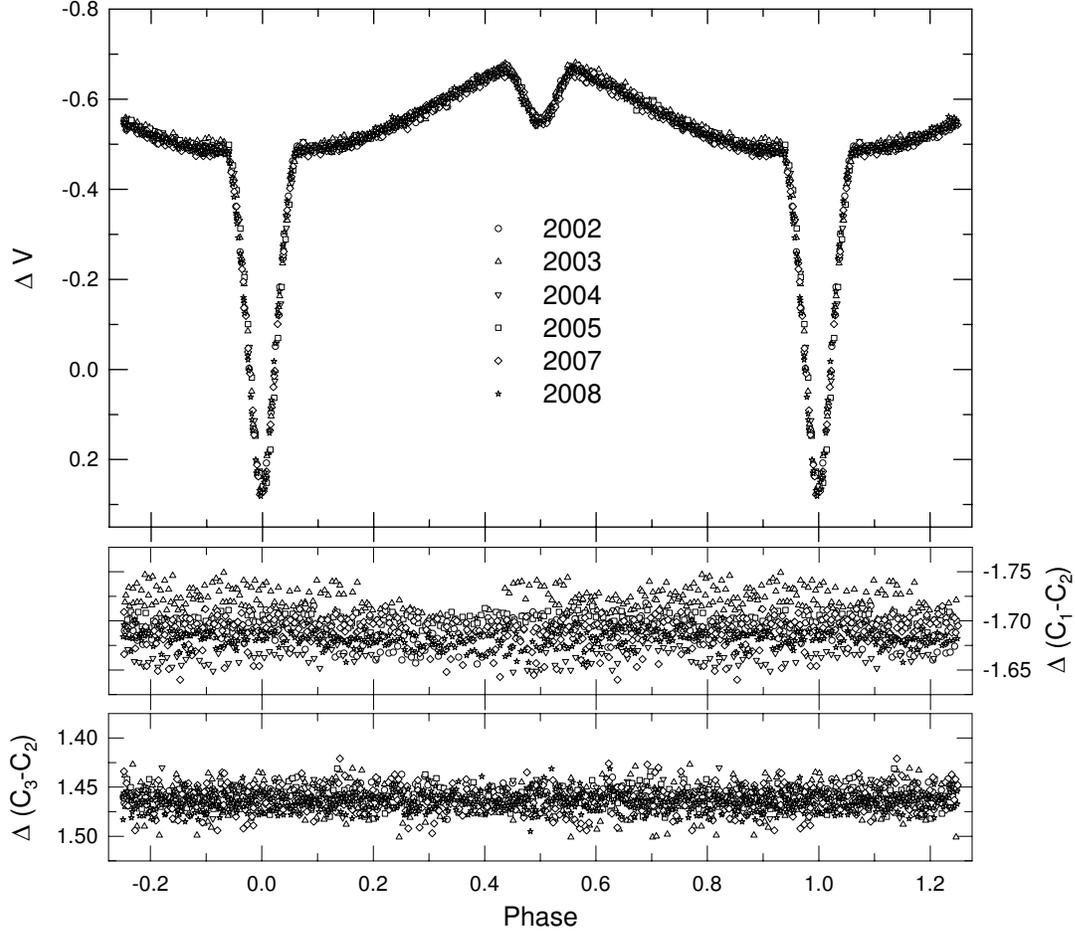}
 \caption{The upper panel displays the $V$ light curve of HW Vir with respect to our comparison star
  (C$_2$). The differences between the C$_1$ and C$_2$ stars are shown in the middle panel and
  the magnitude differences (C$_3$-C$_2$) between the check and second comparison stars in the lower panel.}
 \label{Fig2}
\end{figure}

\begin{figure}
 \includegraphics[]{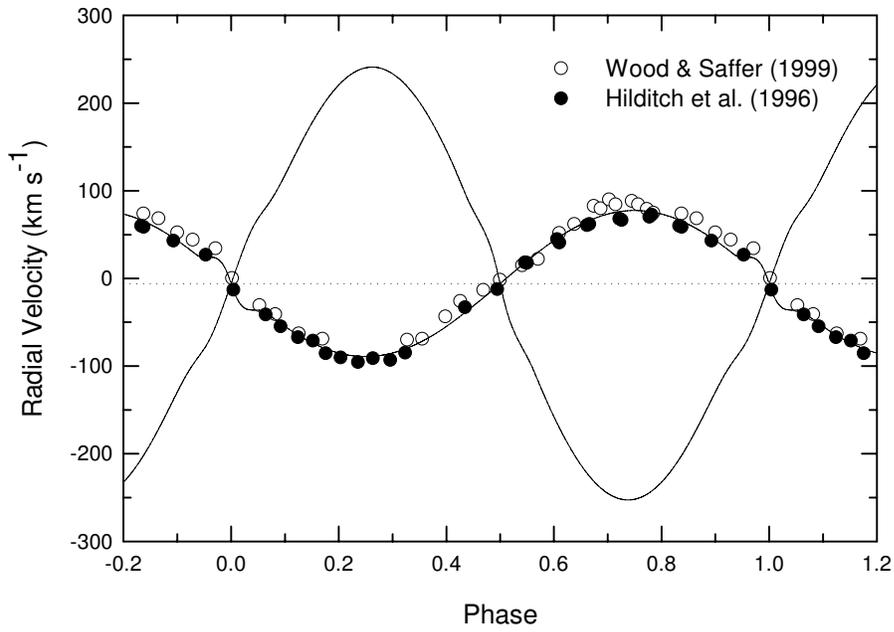}
 \caption{Radial-velocity curves of HW Vir. The open and filled circles are the measures of Wood \& Saffer
 and Hilditch et al., respectively, while the continuous curves denote the result from consistent light
 and velocity curve analysis.}
\label{Fig3}
\end{figure}

\begin{figure}
 \includegraphics[]{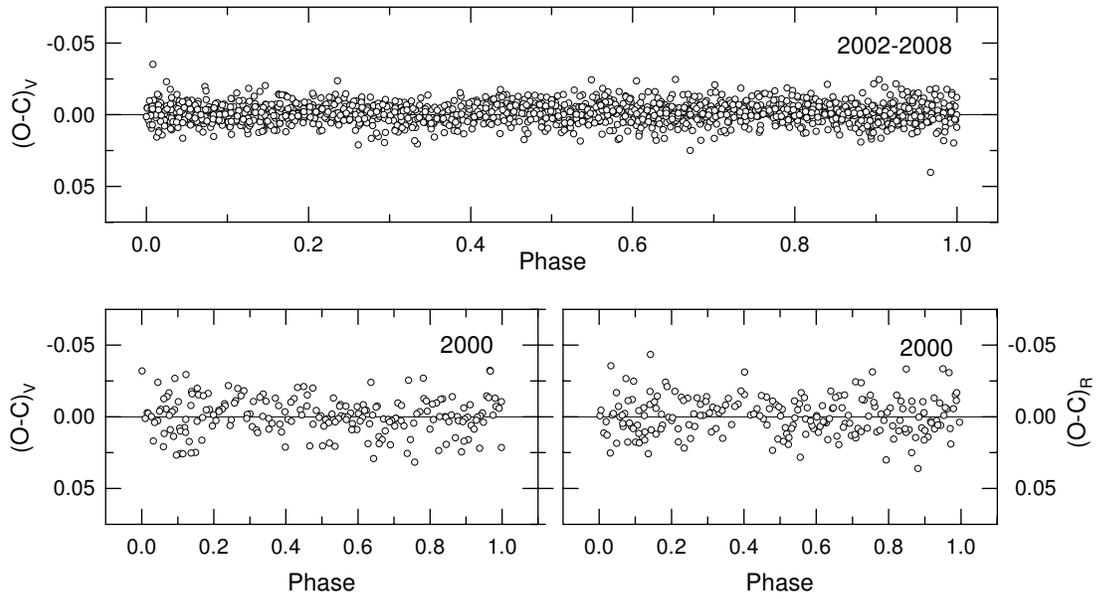}
 \caption{The upper plot shows the residuals of the 2002--2008 $V$ observations from the binary parameters
 in Model 1 of Table 3. The lower two panels represent the residuals in $V$ and $R$ from the solution
 of Table 4 for the 2000 light curves.}
 \label{Fig4}
\end{figure}

\begin{figure}
 \includegraphics[]{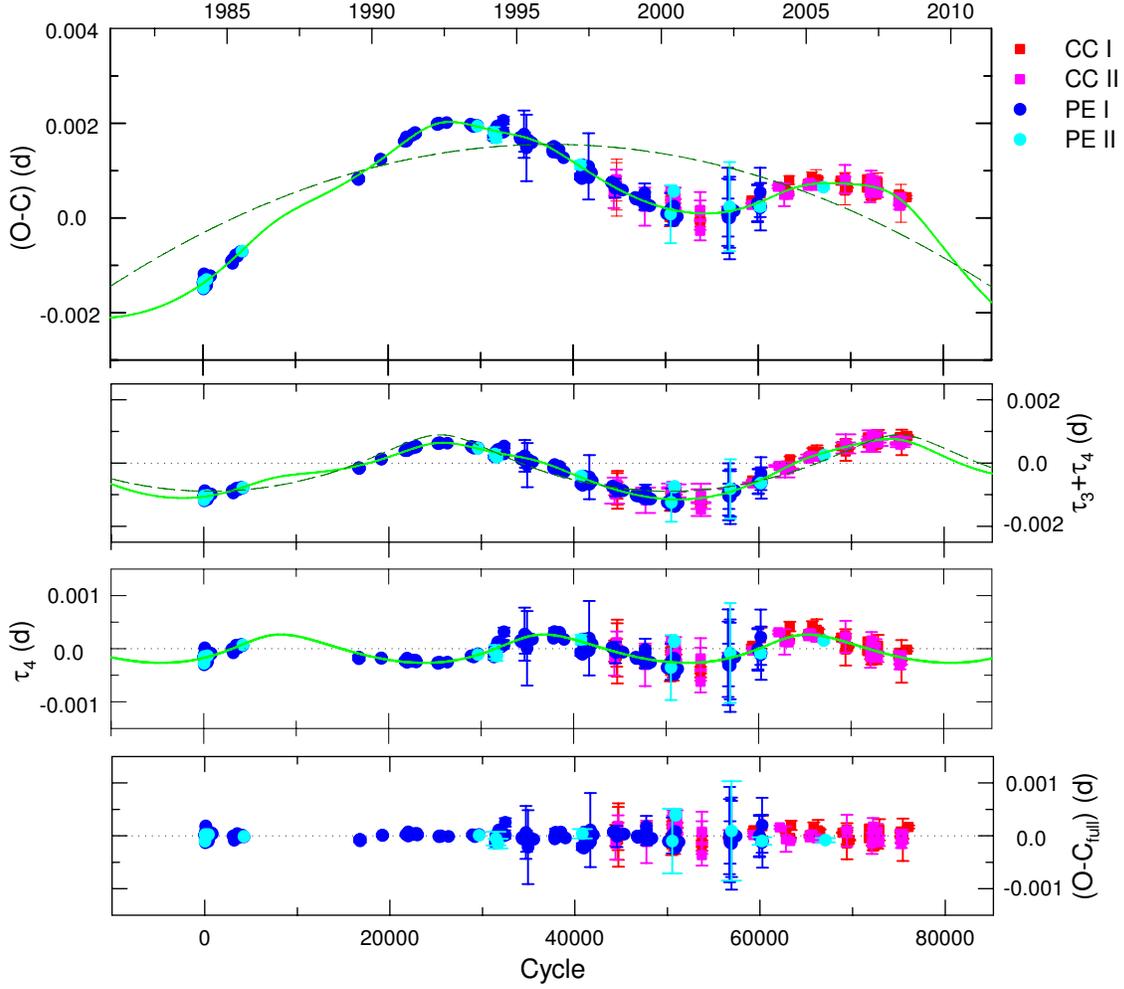}
 \caption{In the top panel the $O$--$C$ diagram of HW Vir constructed with the linear terms of
 the \. Ibanoglu et al. ephemeris.  The quadratic {\it plus} two-LTT ephemeris is drawn as the solid curve and
 the dashed parabola is due to only the quadratic term  of equation (2). The second and third panels display
 the residuals ($\tau_3$+$\tau_4$) from the quadratic term and the residuals $\tau_4$ from the quadratic term
 plus $\tau_3$, respectively. In the second panel, the dashed curve represents the $\tau_3$ light-time orbit.
 The lowest panel shows the residuals from the complete equation. In all panels, error bars are shown for
 only the timings with known errors.}
\label{Fig5}
\end{figure}

\clearpage
\begin{deluxetable}{ccccc}
\tablewidth{0pt} \tablecaption{CCD photometric observations of HW Vir.}
\tablehead{ \colhead{HJD}  & \colhead{$\Delta V$} && \colhead{HJD} & \colhead{$\Delta R$}} \startdata
 2,451,630.13244  &  1.197  &&  2,451,630.13292  &  1.541  \\
 2,451,630.13447  &  1.197  &&  2,451,630.13495  &  1.550  \\
 2,451,630.13649  &  1.209  &&  2,451,630.13697  &  1.550  \\
 2,451,630.13850  &  1.248  &&  2,451,630.13899  &  1.595  \\
 2,451,630.14054  &  1.429  &&  2,451,630.14102  &  1.817  \\
 2,451,630.14257  &  1.728  &&  2,451,630.14304  &  2.123  \\
 2,451,630.14458  &  1.944  &&  2,451,630.14506  &  2.293  \\
 2,451,630.14660  &  1.817  &&  2,451,630.14708  &  2.090  \\
 2,451,630.14862  &  1.543  &&  2,451,630.14911  &  1.802  \\
 2,451,630.15065  &  1.295  &&  2,451,630.15111  &  1.594  \\
\enddata
\tablecomments{This sample is shown for guidance regarding its form and content. The full table is presented
 in its entirety in the electronic edition of the Astronomical Journal.}
\end{deluxetable}

\begin{deluxetable}{lccc}
\tablewidth{0pt}
\tablecaption{Radial velocity and light-curve sets for HW Vir.}
\tablehead{
\colhead{Reference}       & \colhead{Season} & \colhead{Data type} & \colhead{$\sigma$$\rm ^a$} }
\startdata
Wood \& Saffer            & 1993             & RV1                 & 5.19 km s$^{-1}$   \\
Hilditch et al.           & 1994             & RV1                 & 3.39 km s$^{-1}$   \\
SOAO                      & 2000             & $V,R$               & 0.0125             \\
SOAO                      & 2002--2008       & $V$                 & 0.0077             \\
\enddata
\tablenotetext{a}{For the light curves, in units of total light at phase 0.25.}
\end{deluxetable}

\begin{deluxetable}{cccccc}
\tablewidth{0pt}
\tablecaption{Velocity and light curve parameters of HW Vir.}
\tablehead{
\colhead{Parameter}      & \multicolumn{2}{c}{Model 1}                       && \multicolumn{2}{c}{Model 2}                       \\ [1.0mm] \cline{2-3} \cline{5-6} \\ [-2.0ex]
                         & \colhead{Primary}         & \colhead{Secondary}   && \colhead{Primary}         & \colhead{Secondary}
}
\startdata
$\gamma$ (km s$^{-1}$)   & \multicolumn{2}{c}{-5.80$\pm$0.64}                && \multicolumn{2}{c}{-9.07$\pm$0.63}                \\
$a$ (R$_\odot$)          & \multicolumn{2}{c}{0.8594$\pm$0.0089}             && \multicolumn{2}{c}{0.8510$\pm$0.0087}             \\
$q$                      & \multicolumn{2}{c}{0.2931$\pm$0.0043}             && \multicolumn{2}{c}{0.2931$\pm$0.0042}             \\
$i$ (deg)                & \multicolumn{2}{c}{80.98$\pm$0.10}                && \multicolumn{2}{c}{80.98$\pm$0.10}               \\
$T$ (K)                  & 28488$\pm$208$\rm ^{a,b}$ & 3084$\pm$272          && 28488$\pm$208$\rm ^{a,b}$ & 3084$\pm$267          \\
$\Omega$                 & 5.020$\pm$0.035           & 2.806$\pm$0.021       && 5.030$\pm$0.035           & 2.807$\pm$0.021       \\
$\Omega_{\rm in}$        & \multicolumn{2}{c}{2.451}                         && \multicolumn{2}{c}{2.451}                         \\
$A$                      & 1.0$\rm ^b$               & 0.90$\pm$0.60         && 1.0$\rm ^b$               & 0.90$\pm$0.62         \\
$g$                      & 1.0$\rm ^b$               & 0.32$\rm ^b$          && 1.0$\rm ^b$               & 0.32$\rm ^b$          \\
$X$                      & 0.61$\rm ^b$              & 0.27$\rm ^b$          && 0.61$\rm ^b$              & 0.27$\rm ^b$          \\
$x_{V}$                  & 0.24$\pm$0.09             & 0.47$\pm$0.19         && 0.24$\pm$0.09             & 0.47$\pm$0.18         \\
$L/(L_1+L_2)_{V}$        & 0.9997$\pm$0.0008         & 0.0003$\pm$0.0002     && 0.9997$\pm$0.0007         & 0.0003$\pm$0.0002     \\
$r$ (pole)               & 0.2113$\pm$0.0016         & 0.1991$\pm$0.0034     && 0.2108$\pm$0.0016         & 0.1990$\pm$0.0033     \\
$r$ (point)              & 0.2137$\pm$0.0017         & 0.2130$\pm$0.0047     && 0.2132$\pm$0.0017         & 0.2128$\pm$0.0046     \\
$r$ (side)               & 0.2126$\pm$0.0016         & 0.2014$\pm$0.0036     && 0.2121$\pm$0.0016         & 0.2013$\pm$0.0035     \\
$r$ (back)               & 0.2134$\pm$0.0016         & 0.2096$\pm$0.0043     && 0.2129$\pm$0.0016         & 0.2095$\pm$0.0042     \\
$r$ (volume)$\rm ^c$     & 0.2124                    & 0.2036                && 0.2120                    & 0.2035                \\
\enddata
\tablenotetext{a}{Value given by Wood \& Saffer.}
\tablenotetext{b}{Fixed parameter.}
\tablenotetext{c}{Mean volume radius.}
\end{deluxetable}

\begin{deluxetable}{ccc}
\tablewidth{0pt}
\tablecaption{Photometric parameters obtained from the 2000 observations.}
\tablehead{
\colhead{Parameter}      & \colhead{Primary}         & \colhead{Secondary}
}
\startdata
$q$                      & \multicolumn{2}{c}{0.293$\pm$0.014}               \\
$i$ (deg)                & \multicolumn{2}{c}{80.92$\pm$0.36}                \\
$T$ (K)                  & 28488$\pm$208             & 3084$\pm$889          \\
$\Omega$                 & 5.020$\pm$0.108           & 2.806$\pm$0.077       \\
$\Omega_{\rm in}$        & \multicolumn{2}{c}{2.451}                         \\
$A$                      & 1.0                       & 0.90$\pm$1.96         \\
$x_{V}$                  & 0.24$\pm$0.28             & 0.53$\pm$0.57         \\
$x_{R}$                  & 0.21$\pm$0.30             & 0.50$\pm$0.55         \\
$L/(L_1+L_2)_{V}$        & 0.9997$\pm$0.0027         & 0.0003$\pm$0.0007     \\
$L/(L_1+L_2)_{R}$        & 0.9988$\pm$0.0038         & 0.0012$\pm$0.0023     \\
$r$ (pole)               & 0.2113$\pm$0.0049         & 0.1991$\pm$0.0120     \\
$r$ (point)              & 0.2137$\pm$0.0051         & 0.2130$\pm$0.0164     \\
$r$ (side)               & 0.2126$\pm$0.0050         & 0.2014$\pm$0.0125     \\
$r$ (back)               & 0.2134$\pm$0.0051         & 0.2096$\pm$0.0150     \\
$r$ (volume)             & 0.2124                    & 0.2036                \\
\enddata
\end{deluxetable}

\begin{deluxetable}{ccc}
\small
\tablewidth{0pt}
\tablecaption{Absolute dimensions of HW Vir assuming co-planar orbits.}
\tablehead{
\colhead{Parameter}      & \colhead{Primary}        & \colhead{Secondary}}
\startdata
$M$/M$_\odot$            &  0.485$\pm$0.013         &  0.142$\pm$0.004          \\
$R$/R$_\odot$            &  0.183$\pm$0.026         &  0.175$\pm$0.026          \\
$\log$ $g$ (cgs)         &  5.60$\pm$0.12           &  5.10$\pm$0.13            \\
$\log$ $\rho$/$\rho_\odot$  &  1.90$\pm$0.18        &  1.42$\pm$0.19            \\
$L_{\rm bol}$/L$_\odot$  &  19.7$\pm$5.6            &  0.003$\pm$0.001          \\
$M_{\rm bol}$ (mag)      &  1.46$\pm$0.24           &  11.20$\pm$0.46           \\
BC (mag)                 &  -2.76                   &  -4.39                    \\
$M_{V}$ (mag)            &  4.22$\pm$0.24           &  15.59$\pm$0.46           \\
Distance (pc)            &  \multicolumn{2}{c}{181$\pm$20}                      \\
\enddata
\end{deluxetable}

\begin{deluxetable}{llccllcc}
\tablewidth{0pt}
\tablecaption{New times of minimum light for HW Vir.}
\tablehead{
\colhead{HJD} & \colhead{Error} & \colhead{Filter$\rm ^a$} & \colhead{Type} & \colhead{HJD} & \colhead{Error} & \colhead{Filter$\rm ^a$} & \colhead{Type} \\
\colhead{(2,450,000+)} & & & & \colhead{(2,450,000+)} & & & }
\startdata
1,629.26976  &  $\pm$0.00008  & $V,R$  & II   &   4,155.31331  &  $\pm$0.00008  & $V$  & II   \\
1,630.14497  &  $\pm$0.00004  & $V,R$  & I    &   4,155.37148  &  $\pm$0.00001  & $V$  & I    \\
1,630.20327  &  $\pm$0.00021  & $V,R$  & II   &   4,158.23141  &  $\pm$0.00016  & $V$  & II   \\
1,630.26164  &  $\pm$0.00004  & $V,R$  & I    &   4,158.28939  &  $\pm$0.00007  & $V$  & I    \\
1,688.03779  &  $\pm$0.00005  & $V,R$  & I    &   4,158.34788  &  $\pm$0.00005  & $V$  & II   \\
1,688.09631  &  $\pm$0.00012  & $V,R$  & II   &   4,214.08137  &  $\pm$0.00003  & $V$  & I    \\
1,689.02975  &  $\pm$0.00002  & $V,R$  & II   &   4,214.13993  &  $\pm$0.00015  & $V$  & II   \\
1,689.08811  &  $\pm$0.00011  & $V,R$  & I    &   4,509.14824  &  $\pm$0.00003  & $N$  & I    \\
2,342.19202  &  $\pm$0.00009  & $V$    & II   &   4,509.20650  &  $\pm$0.00011  & $N$  & II   \\
2,342.25033  &  $\pm$0.00001  & $V$    & I    &   4,509.26494  &  $\pm$0.00001  & $N$  & I    \\
2,650.33163  &  $\pm$0.00003  & $V$    & II   &   4,512.29956  &  $\pm$0.00001  & $V$  & I    \\
2,650.39008  &  $\pm$0.00002  & $V$    & I    &   4,513.35011  &  $\pm$0.00006  & $V$  & I    \\
2,675.36802  &  $\pm$0.00003  & $V$    & I    &   4,514.16706  &  $\pm$0.00001  & $V$  & I    \\
3,061.30120  &  $\pm$0.00024  & $V$    & II   &   4,514.22530  &  $\pm$0.00008  & $V$  & II   \\
3,061.35970  &  $\pm$0.00003  & $V$    & I    &   4,514.28379  &  $\pm$0.00001  & $V$  & I    \\
3,124.97199  &  $\pm$0.00009  & $V$    & I    &   4,514.34214  &  $\pm$0.00012  & $V$  & II   \\
3,384.32264  &  $\pm$0.00001  & $V$    & I    &   4,515.33427  &  $\pm$0.00003  & $V$  & I    \\
3,384.38102  &  $\pm$0.00005  & $V$    & II   &   4,517.31850  &  $\pm$0.00001  & $V$  & I    \\
3,825.23062  &  $\pm$0.00008  & $N$    & II   &   4,517.37686  &  $\pm$0.00013  & $V$  & II   \\
3,825.28904  &  $\pm$0.00002  & $N$    & I    &   4,607.07585  &  $\pm$0.00003  & $N$  & I    \\
4,155.25477  &  $\pm$0.00001  & $V$    & I    &                &                &      &      \\
\enddata
\tablenotetext{a}{$N$ indicates the eclipse timings made without any filter at CbNUO.}
\end{deluxetable}

\begin{deluxetable}{cccc}
\tablewidth{0pt}
\tablecaption{Parameters for the quadratic {\it plus} two-LTT ephemeris of HW Vir.}
\tablehead{
\colhead{Parameter}            &  Long-term ($\tau_{3}$)            &  Short-term ($\tau_{4}$)        & \colhead{Unit}
}
\startdata
$T_0$                          &  \multicolumn{2}{c}{2,445,730.557123$\pm$0.000025}                   &   HJD            \\
$P$                            &  \multicolumn{2}{c}{0.11671959933$\pm$0.00000000055}                 &   d              \\
$A$                            &  \multicolumn{2}{c}{-(1.323$\pm$0.010)$\times 10^{-12}$}             &   d              \\
$a_{12}\sin i$                 &  0.1551$\pm$0.0054                 &  0.0468$\pm$0.0066              &   AU             \\
$\omega$                       &  90.8$\pm$2.8                      &  60.6$\pm$7.1                   &   deg            \\
$e$                            &  0.46$\pm$0.05                     &  0.31$\pm$0.15                  &                  \\
$n$                            &  0.0622$\pm$0.0005                 &  0.1085$\pm$0.0026              &   deg d$^{-1}$   \\
$T$                            &  2,454,500$\pm$39                  &  2,449,840$\pm$63               &   HJD            \\
$P_{3,4}$                      &  15.84$\pm$0.14                    &  9.08$\pm$0.22                  &   yr             \\
$K_{3,4}$                      &  0.000896$\pm$0.000031             &  0.000267$\pm$0.000038          &   d              \\
$f(M_{3,4})$                   &  (1.486$\pm$0.054)$\times 10^{-5}$ & (1.24$\pm$0.18)$\times 10^{-6}$ &   M$_\odot$      \\
$M_{3,4} \sin i_{3,4}$         &  0.01836$\pm$0.00023               &  0.00809$\pm$0.00040            &   M$_\odot$      \\
$a_{3,4} \sin i_{3,4}$         &  5.30$\pm$0.23                     &  3.62$\pm$0.52                  &   AU             \\[0.5mm]
d$P$/d$t$                      &  \multicolumn{2}{c}{-(8.28$\pm$0.06)$\times 10^{-9}$}                &   d yr$^{-1}$    \\
$\Sigma (O-C)^2$               & \multicolumn{2}{c}{0.00010}                                          &   d              \\
\enddata
\end{deluxetable}

\begin{deluxetable}{cccc}
\tablewidth{0pt}
\tablecaption{Model parameters for the magnetic activity of the cool secondary of HW Vir.}
\tablehead{
\colhead{Parameter}       & Long-term ($\tau_{3}$)   & Short-term ($\tau_{4}$)  & Unit
}
\startdata
$\Delta P$                & 0.0099                   & 0.0052                   &  s                  \\
$\Delta P/P$              & ${9.78\times10^{-7}}$    & ${5.11\times10^{-7}}$    &                     \\
$\Delta Q$                & ${1.10\times10^{47}}$    & ${5.74\times10^{46}}$    &  g cm$^2$           \\
$\Delta J$                & ${5.52\times10^{45}}$    & ${2.88\times10^{45}}$    &  gcm$^{2}$ s$^{-1}$ \\
$I_{\rm s}$               & ${2.79\times10^{51}}$    & ${2.79\times10^{51}}$    &  gcm$^{2}$          \\
$\Delta \Omega$           & ${1.97\times10^{-6}}$    & ${1.03\times10^{-6}}$    &  s$^{-1}$           \\
$\Delta \Omega / \Omega$  & 0.0032                   & 0.0017                   &                     \\
$\Delta E$                & ${2.18\times10^{40}}$    & ${5.95\times10^{39}}$    &  erg                \\
$\Delta L_{\rm rms}$      & ${1.37\times10^{32}}$    & ${6.51\times10^{31}}$    &  erg s$^{-1}$       \\
                          & 0.0351                   & 0.0167                   &  L$_\odot$          \\
                          & 11.7114                  & 5.5668                   &  L$_{2}$            \\
                          & $\pm$0.0019              & $\pm$0.0009              &  mag                \\
$B$                       & 34736                    & 33133                    &  G                  \\
\enddata
\end{deluxetable}

\end{document}